\documentclass[%
 aip,
 amsmath,amssymb,
 reprint,%
]{revtex4-1}

\usepackage{graphicx}% Include figure files
\usepackage{dcolumn}% Align table columns on decimal point
\usepackage{bm}% bold math
\usepackage[utf8]{inputenc}
\usepackage[T1]{fontenc}
\usepackage{mathptmx}
\usepackage{etoolbox}
\usepackage{comment}

\usepackage{mathptmx}

\newcommand{\acrolein}{\textbf{I}}
\newcommand{\pNA}{\textbf{II}}
\newcommand{\quinolinium}{\textbf{III}}
\newcommand{\phtalimide}{\textbf{IV}}
\newcommand{\bimane}{\textbf{V}}
\newcommand{\cyanine}{\textbf{VI}}
\newcommand{\doxo}{\textbf{VII}}
\newcommand{\betaine}{\textbf{VIII}}
\newcommand{\bodipy}{\textbf{IX}}
\newcommand{\merocyanine}{\textbf{X}}
\newcommand{\citidina}{\textbf{XI}}
\newcommand{\SM}{SM}

\makeatletter
\def\@email#1#2{%
 \endgroup
 \patchcmd{\titleblock@produce}
  {\frontmatter@RRAPformat}
  {\frontmatter@RRAPformat{\produce@RRAP{*#1\href{mailto:#2}{#2}}}\frontmatter@RRAPformat}
  {}{}
}%
\makeatother
\begin{document}

\preprint{AIP/123-QED}

\title[QM/MM Approaches for Solvatochromic Shifts]{Assessing the Quality of QM/MM Approaches to Describe Vacuo-to-water Solvatochromic Shifts}
% Force line breaks with \\
\author{Luca Nicoli}
\author{Tommaso Giovannini}%
 \email{tommaso.giovannini@sns.it}
\author{Chiara Cappelli}
 \email{chiara.cappelli@sns.it}
\affiliation{Scuola Normale Superiore,
	Piazza dei Cavalieri 7, 56126 Pisa, Italy.}%

\date{\today}% It is always \today, today,

\begin{abstract}
The performance of different Quantum Mechanics/Molecular Mechanics embedding models to compute vacuo-to-water solvatochromic shifts are investigated. In particular, both non-polarizable and polarizable approaches are analyzed and computed results as compared to reference experimental data. We show that none of the approaches outperforms the others and that errors strongly depend on the nature of the molecular transition. Thus, we prove that the best choice of embedding model highly depends on the molecular system, and that the use a specific approach as a black-box can lead to significant errors and sometimes totally wrong predictions. 
\end{abstract}

\maketitle

\section{Introduction}

Focused models have a long standing tradition in computational chemistry for the simulation of spectral properties of complex systems. \cite{warshel1976theoretical,senn2009qm,giovannini2020molecular,cappelli2016integrated} Among them, quantum mechanics/molecular mechanics (QM/MM) approaches have become very popular, \cite{warshel1976theoretical,gao1997energy,lin2007qm,senn2009qm,mennucci2019multiscale} due to their strengths in dealing with many diverse external environments, ranging from strongly interacting solvents\cite{giovannini2020molecular} to biomolecular environments.\cite{bondanza2020polarizable,curutchet2017quantum,cupellini2019electronic} Indeed, the increasing popularity of QM/MM is linked to their ability to describe target/environment interactions with an atomistic detail.\cite{senn2009qm,boulanger2018qm}

When applied to solvated systems, the most common QM/MM partition consists of treating the solute at the QM level, and the solvent in terms of classical MM force fields. For a given QM level, the quality of QM/MM results strongly depends on the physics lying behind the specific approach which is exploited to model the interaction between the QM and MM layers.\cite{mennucci2019multiscale} The latter is generally limited to electrostatic terms, being non-electrostatic contributions only rarely taken into account.\cite{giovannini2017general,giovannini2019effective,giovannini2019quantum,olsen2015polarizable}%REF!!!! 

The MM layer can be modeled in terms of a set of fixed multipoles placed at atomic sites, thus yielding the so-called Electrostatic Embedding (EE) approach. \cite{senn2009qm} As a consequence, the MM layer polarizes the QM density, but not viceversa. However, a correct physical description of an interacting solute-solvent systems requires mutual solute-solvent polarization effects to be considered.\cite{jensen2003discreteDFT,boulanger2012solvent,boulanger2014toward}
Thus, many different polarizable embedding have been proposed and amply tested. \cite{jensen2003discreteDFT,loco2016qm,olsen2011molecular,boulanger2012solvent,boulanger2014toward,giovannini2019polarizable,bondanza2020polarizable,wu2017simulating,giovannini2020molecular,loco2021atomistic,bondanza2020polarizable}%REFS

%TOMMASO A CHIARA: non ho saputo inserire cit a jacquemin ... 

In polarizable QM/MM approaches, MM fragments are endowed with polarizable multipolar charge distributions which are modified as a result of the interaction with the QM density, and viceversa.\cite{mennucci2019multiscale} The physically consistent description which is then obtained, permits to compute remarkably accurate values of many spectroscopic signals, especially when polarizable QM/MM approaches are coupled to accurate procedures to sample the configurational phase-space. \cite{jensen2005first,caricato2013vertical,giovannini2019simulating,segatta2019quantum,nottoli2021multiscale} 
The various QM/MM approaches differ from the specific way the electrostatic and polarization terms are modeled. % the MM portion, which can be grounded in an electrostatic multipolar expansion or in conceptual Density Functional Theory (DFT).\cite{stone2013theory,geerlings2003conceptual} 
The latter not only modifies the solute's ground state density, but also its response properties.

Despite the increasing interest in exploiting QM/MM approaches to describe spectral properties, the performance of the different QM/MM approaches has only rarely been investigated.\cite{reinholdt2018modeling, prioli2021modeling} Therefore, the ideal model to be employed for a given application has not been clearly defined yet.

In this work, we present extensive comparison of the results obtained by applying a selection of QM/MM embedding models to the calculation of vacuo-to-water solvatochromic shifts. The approaches are chosen because they conceptually span diverse classes of models that are employed in the literature. In particular, we employ the EE (as specified by means of the TIP3P parametrization), \cite{jorgensen1983comparison} where MM atoms are described in terms of fixed charges, the polarizable Fluctuating Charges (FQ), \cite{rick1994dynamical,chen2009charge,chen2008unified,giovannini2020molecular} where polarization effects are described in terms of a set of charges that vary as a response to the external electric potential.\cite{cappelli2016integrated} Discrete Reaction Field (DRF) \cite{jensen2003discreteDFT,jensen2003discreteResponse} is an example of amply used approaches which model polarization effects in terms of a set of induced dipoles assigned to MM atoms.\cite{jensen2003discreteDFT,jensen2003discreteResponse,olsen2011molecular,curutchet2009electronic,list2016excited} More sophisticated models are used to refine DRF electrostatic description in terms of fixed multipolar expansions. \cite{loco2016qm,ren2002consistent,ren2003polarizable} The last approach is the Fluctuating Charges and Fluctuating Dipoles (FQF$\mu$) model,\cite{giovannini2019polarizable} where each MM atom is assigned a charge and dipole which can vary as a result of polarization effects. While EE and DRF directly follow from an electrostatic multipolar expansion of the energy,\cite{stone2013theory} FQ is grounded in conceptual DFT,\cite{geerlings2003conceptual} and FQF$\mu$ can be seen as a pragmatical extension of FQ.\cite{giovannini2019polarizable}

%In fact, \cite{jorgensen1983comparison} in FQ model, In the DRF approach, each MM atom is endowed with fixed charges and induced dipoles.\cite{jensen2003discreteDFT,jensen2003discreteResponse} It is worth noting that, in our work,  Finally, in FQF$\mu$ both charges and dipoles can respond to the external electric potential/field.\cite{giovannini2019polarizable,giovannini2019electronic} 

Each embedding approach models QM/MM interactions according to the order of the multipolar expansion of the MM variables (fixed and/or polarizable). From the numerical point of view, such an interaction also depends on the parameters defining the specific model: fixed atomic charge $q$ (for EE and DRF), atomic electronegativity $\chi$ and chemical hardness $\eta$ (for FQ and FQF$\mu$), and atomic polarizability $\alpha$ (for DRF and FQF$\mu$). The numerical values of such parameters clearly determine the QM/MM interaction, and in turn computed spectroscopic signals.\cite{ambrosetti_quantum_2021} Thus, in this paper a total of eight different parameter sets, which are specifically developed for the aqueous environment, are compared. \cite{jorgensen1983comparison,rick1994dynamical,carnimeo2015analytical,giovannini2019effective,jensen2003discreteDFT,giovannini2019polarizable}  

%The shifts calculated by each embedding model are then compared with available and fitted experimental shifts, in order to have a physically sound reference scale. 

\noindent
%Our results show that it is not possible to find an embedding model that invariably performs better than the others since the performance of each model is highly system dependent. 

% Nevertheless, our analysis allows to outline some rule of thumbs which can be helpful to select the most suitable QM/MM embedding model for calculating vertical transition energies of a target molecule in water.

\noindent
The manuscript is organized as follows. In the next section, we briefly recap the theoretical foundations of the QM/MM embedding approaches which are exploited in this work. Then, their performance are tested to describe vacuo-to-water solvatochromic shift of a set of 11 medium-to-large molecules, for which experimental data are available in the literature. Results are also discussed in terms of the physico-chemical description of the QM/MM interaction and the nature of the solute's transition.

%in the following section we briefly describe the theoretical foundations which are common to all the QM/MM embedding approaches treated in this paper. In this framework we define the set of characteristics operators which define the working equations both in ground state and for the determination of vertical transition energies.
%In \cref{sec:models}, we present the specifics of the different embedding models, in terms of the operators defined in \cref{sec:theory}. In \cref{sec:computational details} we  discuss the computational details of the work and eventually in \cref{sec:results} we discuss the performance of the models in terms of statistical estimators. We rationalize the behaviour of the different models in terms of the physico-chemical description they provide and in terms of a simplified two-
%states model for the blue-shifting transitions.

\section{Theoretical Modelling}\label{sec:theory}
The total energy of a QM/MM system reads: \cite{giovannini2020molecular}
\begin{align}
    \label{eq:Etot}
    E^{tot} & = E_{QM} + E_{MM} + E_{QM/MM}^{int} 
\end{align}
where $E_{QM}$ and $E_{MM}$ are the energies of the QM and MM portions, respectively. By neglecting non-electrostatic (dispersion/repulsion) interactions, the QM-MM interaction energy $E^{int}_{QM/MM}$ can be expressed as:
\begin{align}
    \label{eq:Eint}
    E_{QM/MM}^{int} & = E_{QM/MM}^{ele} + E_{QM/MM}^{pol}
\end{align}
where the electrostatic $E_{QM/MM}^{ele}$ and possibly polarization $E_{QM/MM}^{pol}$ energy terms are highlighted. In a generic definition of a force field, MM atoms can be endowed with a fixed multipolar distribution $\mathbf{M}$ (charges, dipoles, quadrupoles, ...) and additional quantities $\mathbf{D}$, accounting for polarization effects. By this, the various polarizable or non-polarizable QM/MM approaches differ in the way they define $\mathbf{M}$ and $\mathbf{D}$, and because they account or neglect polarization terms (i.e. $\mathbf{D}$).
%For sake of clarity $\mathbf{D}^{\dagger} = \left[\mathbf{D}^{0}, \mathbf{D}^1, \dots \mathbf{D}^{N}\right]$, where $\mathbf{D}^{0}$ is a set of charges $\mathbf{q}$, $\mathbf{D}^{1}$ is a set of dipoles and so on. Analogous definitions can be done for $\textbf{M}^{\dagger}$. Interaction between the different moieties happen by means of classical Coulomb interaction. 
%
By assuming a classical electrostatic interaction between the QM and MM portions, the total energy in Eq. \ref{eq:Etot} can be rewritten as:
\begin{equation}
\label{eq:functional form total energy}
\begin{aligned}
   \mathcal{E}^{tot}[\rho, \mathbf{D}] =&   E_{QM}[\rho(\mathbf{r})] +\mathbf{M}^{\dagger} \int\mathbf{T}_M(\mathbf{r})\rho(\mathbf{r})d\mathbf{r} \\
    & + \frac{1}{2}  \mathbf{D}^{\dagger}\mathbf{A} \mathbf{D}  + \mathbf{D}^{\dagger}\int\mathbf{T}_D(\mathbf{r})\rho(\mathbf{r})d\mathbf{r} + \mathbf{D}^{\dagger}\mathbf{T}\mathbf{M}
    \end{aligned}
\end{equation}
where the $\mathbf{A}$ matrix describes the self interaction of the polarization sources; $\mathbf{T}$ is a block matrix, which takes into account the interaction between the fixed and polarizable MM distributions. $\mathbf{T}_\xi(\mathbf{r})$ ($\xi = M, D$) collects QM/MM electrostatic interaction kernels \cite{stone2013theory} (see Sec. S1.1 in the Supplementary Material -- SM for more details).

\noindent
Within a Kohn–Sham (KS) density functional theory (DFT) formulation, by differentiating Eq. \ref{eq:functional form total energy} with respect to $\rho$, the QM/MM Fock Matrix $\Tilde{F}$ is recovered. By minimizing Eq. \ref{eq:functional form total energy} with respect to $\mathbf{D}$, the equations which describe the polarization of the MM portion are obtained. This allows us to define the coupled QM/MM equations:

\begin{align}
    &\label{eq:QMcoupled} \frac{\delta\mathcal{E}^{tot}[\rho,\mathbf{D}]}{\delta \rho(\mathbf{r})} = h_{KS}^0[\rho(\mathbf{r})] + \hat{v}^{emb}(\mathbf{r}) = \Tilde{F}\\
    &\label{eq:MMcoupled}  \frac{\delta\mathcal{E}^{tot}[{\rho},\mathbf{D}]}{\delta \mathbf{D}} = \bm{\Theta} [{\rho},\mathbf{D}] = 0
\end{align}

where $h^0_{KS}$ is the common KS operator, given by:

\begin{equation}
\begin{aligned}
h_{\mathrm{KS}}^{0} &= -\frac{1}{2} \nabla^{2}-\sum_{m} \frac{Z_{m}}{\left|\mathbf{r}-\mathbf{R}_{m}\right|}+\int \frac{\rho(\mathbf{r}')}{\left|\mathbf{r}-\mathbf{r}^{\prime}\right|} d \mathbf{r}^{\prime}+\frac{\delta E_{\mathrm{XC}}}{\delta \rho(\mathbf{r})}
\end{aligned}
\end{equation}

where $E_{XC}$ is the exchange-correlation energy functional. In Eqs. \ref{eq:QMcoupled} and \ref{eq:MMcoupled}, $\hat{v}^{emb}(\mathbf{r})$ and $\bm{\Theta}[\rho,\mathbf{D}]$ are defined as:
\begin{align}
\hat{v}^{emb}(\mathbf{r}) &=\mathbf{M}^{\dagger}\mathbf{T}_M(\mathbf{r}) + \mathbf{D}^{\dagger}\mathbf{T}_D(\mathbf{r})\label{eq:Vembgeneral}\\
\bm{\Theta}[\rho,\mathbf{D}] &= \mathbf{A}\mathbf{D} + \int\mathbf{T}_D(\mathbf{r})\rho(\mathbf{r})d\mathbf{r} + \mathbf{T}\mathbf{M} \label{eq:Thetageneral}
\end{align}

The solutions of Eqs. \ref{eq:QMcoupled} and \ref{eq:MMcoupled} define the ground state (GS) QM density and the polarization vector $\mathbf{D}$. 

Vertical excitation energies can be computed by resorting to the linear response (LR) formulation of the time-dependent DFT (TDDFT) formalism (see Sec. S1 in the \SM~ for more details). \cite{olsen2011molecular,giovannini2020molecular,jensen2003discreteResponse,giovannini2019electronic}. In the case of QM/MM approaches, LR-TDDFT equations \cite{casida_molecular_1998} are modified to account for the presence of the MM layer.\cite{giovannini2019electronic} In particular, the MM environment modifies excitation energies through two mechanisms: (i) modification of energy and spatial distribution of GS molecular orbitals (MOs), usually referred to as indirect effect; (ii) inclusion of additional terms in LR-TDDFT equations, which account for the mutual interaction between the MM layer and the transition QM density. The latter contribution is usually called ``direct effect'', and is only present in case of polarizable embedding approaches. Note that state-specific formulations of polarizable embedding models have also been proposed. They specifically account for the relaxation of the solute density in the excited state of interest, while discarding the dynamical aspects of solute-solvent interactions, which are instead considered in the LR formalism.\cite{giovannini2019electronic,caricato2006formation}
It is finally worth noting that local field effects induced on the QM moiety due to the polarization of the MM portion to the external radiation field, are not taken into account in this work, although they may affect computed oscillator strengths.\cite{list2016local}

\subsection{Embedding Models}\label{sec:models}

The equations reported in the previous section are general enough to constitute a unified framework, which can be specified for the various embedding approaches that are exploited in the present work. The latter differ in the way $\mathbf{D}$, $\mathbf{M}$ are defined.
%LUCA QUI!
%The equations reported in the previous section are general enough to constitute a unified framework for the various embedding approaches that are exploited in the present work, which differ in the way $\mathbf{D}$, $\mathbf{M}$ are defined.
%, and $\mathbf{A}$.
%This in turn depends upon the approximations that are carried out when describing the physical problem. 
%In this section we provide a brief description of the QM/MM approaches used in this work, which range from electrostatic embedding to polarizable embedding. The methods differ for the MM variables defining the electrostatic/polarization energy terms in Eq. \ref{eq:Eint}. In particular, we have considered a hierarchy of embedding models: 

\begin{enumerate}
    \item Electrostatic embedding (EE): each atom in the MM region is endowed with a fixed charge i.e. $\mathbf{M} = [\mathbf{q}_{M}]$ and $\mathbf{D} = [\mathbf{0}]$. Therefore, the MM layer polarizes the QM density but not viceversa, thus it only indirectly affects the QM solute's response properties. %For water molecule, we exploit the TIP3P force field. 
    \item Fluctuating Charges (FQ) approach: each MM atom is endowed with a charge, whose value is not fixed, but varies as a result polarization effects. \cite{rick1995fluctuating,cappelli2016integrated,giovannini2020molecular,giovannini2020theory} Thus, $\mathbf{M} = [\mathbf{0}]$ and $\mathbf{D} = \left[\mathbf{q}\right]$, with $\mathbf{q}$ being the polarizable charges. The parameters entering the FQ models, thus determining the $\mathbf{q}$ charges, are the atomic electronegativity $\chi$ and chemical hardness $\eta$, which are theoretically defined in conceptual DFT.\cite{geerlings2003conceptual} Polarization follows from the electronegativity equalization principle, \cite{mortier1985electronegativity,sanderson1951interpretation} which allows to define atomic partial charges in terms of the constrained minimum of a suitable energy functional. \cite{giovannini2020molecular} More details on the FQ model can be found in section S2 in the {\SM}.
    \item Discrete Reaction Field (DRF): each MM atom is endowed with a fixed charge $q$ and a polarizable dipole $\bm{\mu}$, \cite{jensen2003discreteDFT,jensen2003discreteResponse} This approach to model polarization effects is exploited also by other polarizable QM/MM approaches.\cite{olsen2011molecular,boulanger2012solvent,mennucci2019multiscale,ren2003polarizable} Thus, in this case, $\mathbf{M} = \mathbf{q}_{M}$ and $\mathbf{D} = [ \mathbf{0}, \bm{\mu}]$.  Additional details about DRF can be found in section S3 in the {\SM}.
    \item Fluctuating Charges and Fluctuating Dipoles ({FQF$\mu$}): each MM atom is endowed both with a polarizable charge $q$ and a polarizable dipole $\bm{\mu}$.  \cite{giovannini2019polarizable,giovannini2019electronic,giovannini2019calculation,giovannini2019quantum,marrazzini2020calculation} FQF$\mu$ is a pragmatic extension of the FQ model, where  $\mathbf{D} = \left[\mathbf{q},\bm{\mu}\right]$. The parameters that need to be set are the atomic electronegativity $\chi$, chemical hardness $\eta$ and atomic polarizability $\alpha$. Additional information on FQF$\mu$ is reported in section S4 in the {\SM}.
\end{enumerate}

\begin{figure}[!htb]
    \centering
    \includegraphics[width=0.45\textwidth]{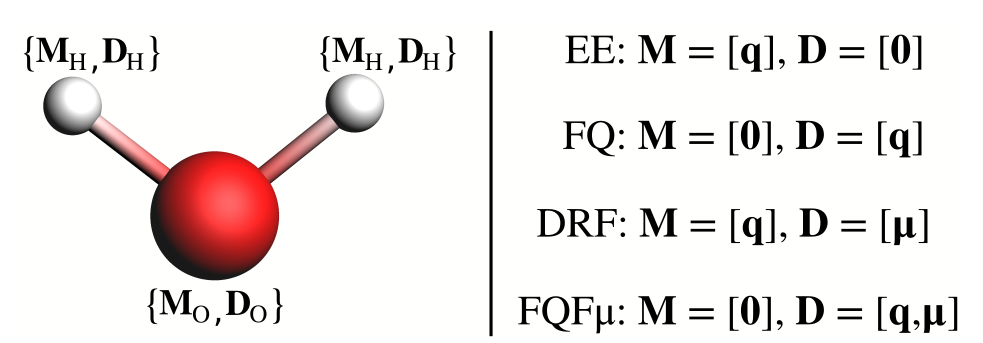}
    \caption{Graphical explanation of the variables associated to EE, FQ, DRF and FQF$\mu$ FFs for the water molecule.}
    \label{fig:wat-models}
\end{figure}

To better understand analogies and differences between the aforementioned approaches, they are schematically specified for the water molecule in Fig. \ref{fig:wat-models}.

\section{Computational details}\label{sec:computational details}

\begin{figure}[!h]
    \centering
    \includegraphics[width=0.5\textwidth]{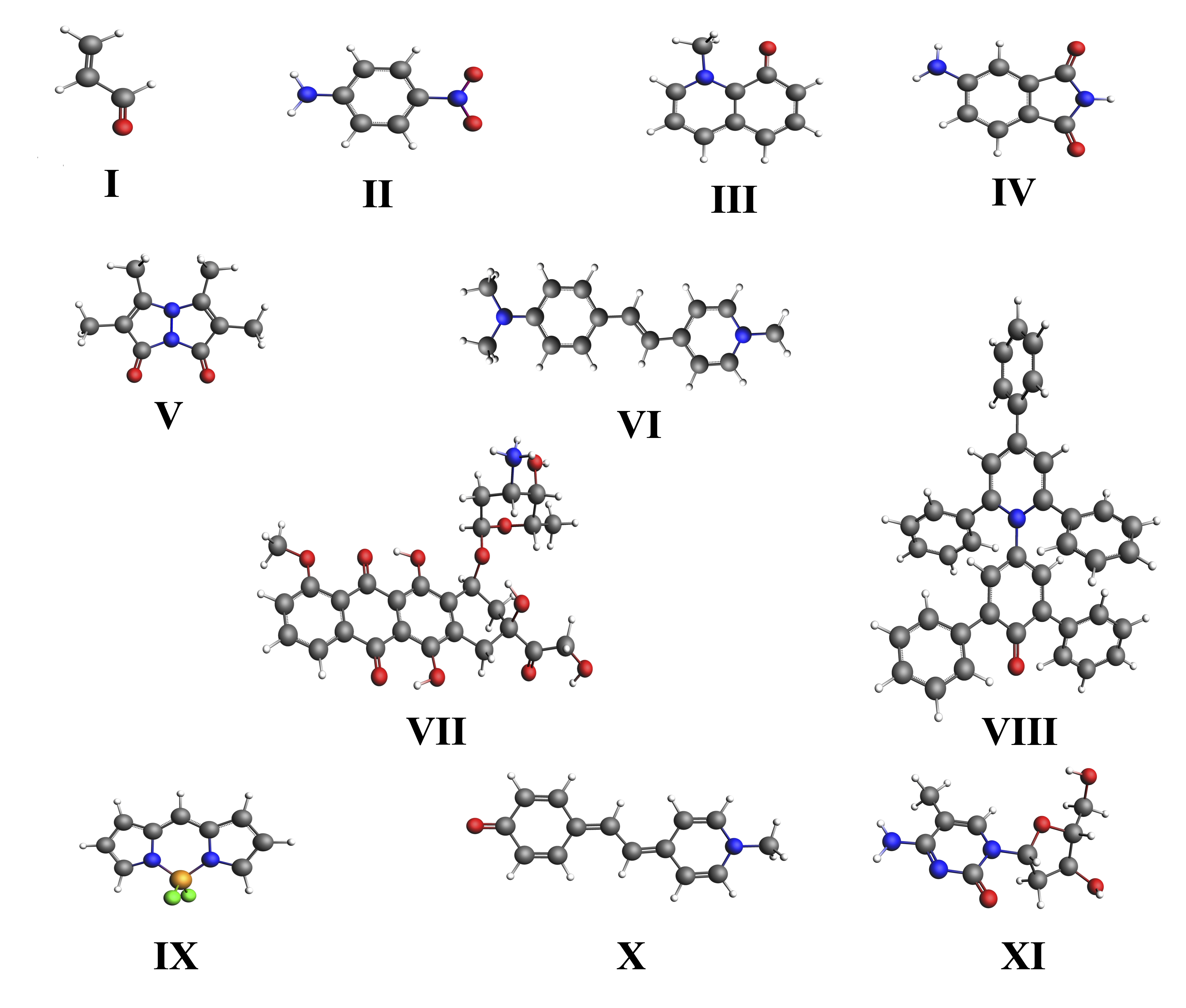}
    \caption{Molecules studied in this work. {\acrolein} acrolein, {\pNA} para-nitroaniline,  {\quinolinium} 1-methyl-8-oxyquinolinium betaine, {\phtalimide} 4-aminophtalimide, {\bimane} syn-$CH_3$-$CH_3$ bimane, {\cyanine} 4-2-[4-(dimethylamino)phenyl]ethenyl-1-methylpyridinium, {\doxo}  doxorubicin, {\betaine}  Reichardt's betaine, {\bodipy}  4,4-difluoro-4-bora-3a,4a-diaza-s-indacene (BODIPY), {\merocyanine}  1-methyl-4-[(oxocyclohexadienylidene)ethylidene]-1,4-dihydropyridine  or Brooker's merocyanine (MB), {\citidina} 5-methylcytidine.}
    \label{fig:molecules set}
\end{figure}

We apply the aforementioned QM/MM approaches to the calculation of vacuo-to-water solvatochromic shifts. To this end, we select eleven molecules (see Fig. \ref{fig:molecules set}) for which experimental UV-Vis absorption spectra in aqueous solution are available in the literature.\cite{aidas2008performance,MILLEFIORI197721,KOVALENKO2000312,novaki1996solvatochromism,soujanya1992nature,politzer1989effects,abdel2009solvatochromism,zakerhamidi2014photo,reichardt2011solvents,arroyo2009smallest,abdel2005absorption,doi:10.1021/j100280a012,martinez2016computing} The variety of molecular size, together with the different sign of experimentally measured solvatochromic shifts, makes this set an ideal test-bed for embedding models.

% In order to account for the rearrangement of the solvent around the solute, the solvent phase-space is sampled by means of molecular dynamics simulations (MD), where the solute is kept frozen at its optimized geometry obtained in vacuo. 
In order to sample the solute-solvent phase-space, molecular dynamics (MD) simulations are performed. In particular, we run MD simulations of {\acrolein}, {\pNA}, {\phtalimide}, {\bimane}, {\cyanine}, {\doxo}, {\bodipy}, and {\citidina} in aqueous solution without imposing any constraints on the solute's geometry. On the other hand, the solute geometry is kept frozen at the PCM \cite{tomasi2005quantum} optimized structure during MD runs of {\quinolinium}, {\betaine}, and {\merocyanine}, because only minor geometrical distortions are expected, due to their limited flexibility. All MDs are performed according to the protocols previously reported by some of the present authors (see Refs. \citenum{giovannini2019quantum}, \citenum{giovannini2019simulating},
\citenum{ambrosetti_quantum_2021},
\citenum{giovannini2019electronic}).
For each molecule, a series of uncorrelated snapshots are extracted from MD runs, and for each snapshot, a spherical droplet with a variable radius depending on the solute intrinsic size is cut. The droplet radius ranges from 15 \AA~ (molecule {\acrolein}) to 25 \AA~ (molecule {\betaine}), and it is set to retain all solute-solvent interactions (see Tab. \ref{tab:geometries} for the average number of water molecules included for each system). For each snapshot, the absorption spectrum is calculated and convoluted with a Gaussian function of FWHM of 0.3 eV. The final UV/Vis absorption spectrum, is then obtained as the average over the set of uncorrelated snapshots extracted from MD runs. Note that the convergence of the computed spectra as a function of the number of snapshots (see Tab. \ref{tab:geometries}) has been checked.
Similarly to previous studies,\cite{aidas2008performance,giovannini2019quantum,giovannini2019electronic} for each investigated molecule we perform an extra set of calculations in which all water molecules that are placed at a distance lower than 3.5 {\AA} from each solute atom are described at the QM level, whereas the remaining ones are treated by using the FQF$\mu$ force field. The resulting approach is called QM/QMw/FQF$\mu$ (QMw). Note that, within this approach, a proper QM description of hydrogen bonding interactions is introduced. QM/QMw/FQF$\mu$ results are obtained as an average on the minimal number of geometries (70 structures for {\acrolein}, 60 structures for {\pNA}, and 20 structures for the remaining molecules) which guarantee the convergence of the spectra (see Section S5.1 in the {\SM}).
% In addition to QM/MM calculations, for each investigated molecule an extra set of calculations was included, which we call QM/QMw/FQF$\mu$ (in the following called QMw). These sets of calculations were performed carving out a spherical droplet of water of 3.5 {\AA} of radius around the solute molecule. These water molecules were treated at the same QM level of theory of the solute, whereas the remaining water molecules were treated using the FQF$\mu$ embedding model. This approach is based on previous studies from some of the present authors (see Ref. \citenum{giovannini2019quantum}). For each molecule such QM/QMw/FQF$\mu$ calculations were performed on a smaller number of geometries compared to pure QM/MM calculations (70 structures for {\acrolein}, 60 structures for {\pNA}, and 20 structures for the remaining molecules). The study of the convergence of the QM/QMw/FQF$\mu$ excitation energy for each molecule and transition in function of the geometries considered is reported in section SXX of the {\SM}.  

For each system, the solvatochromic shift ($\Delta E$) is calculated as:
\begin{equation}
    \Delta E = E_{VAC} - E^{max}_{WAT}
\end{equation}
where $E_{VAC}$ is the excitation energy in gas-phase and $E^{max}_{WAT}$ is the energy of the maximum of the absorption band computed in aqueous solution.
Moreover, in order to classify and to deeply investigate the nature of the electronic transitions, we calculate the $\Delta r$ charge transfer (CT) index by \textit{Guido et al.} \cite{guido2013metric,guido2014effective}. This index is defined in terms of the MOs involved in the electronic transition.

\begin{table}[!htbp]
    \centering
    \begin{ruledtabular}
     \resizebox{.45\textwidth}{!}{  
    \begin{tabular}{l|lllllllllll}
        & {\acrolein}\cite{giovannini2019quantum} & {\pNA}\cite{giovannini2019electronic} & {\quinolinium}\cite{ambrosetti_quantum_2021} & {\phtalimide}\cite{giovannini2019simulating} &{\bimane}\cite{giovannini2019simulating} &  {\cyanine}\cite{giovannini2019simulating} & {\doxo}\cite{giovannini2019simulating} & 
        {\betaine}\cite{ambrosetti_quantum_2021}  & {\bodipy}\cite{giovannini2019simulating} & {\merocyanine}\cite{ambrosetti_quantum_2021}  & {\citidina}\cite{giovannini2019simulating}\\
        \hline
        N$_{\mathrm{frames}}$  &  200 & 100 & 100 & 200 & 200 & 200 & 200 & 100 & 200 & 100 & 200\\
        N$_{\mathrm{H_2O}}$ & 492 & 491 & 868 & 725 & 729 & 1179 & 1208 & 2249 & 727 & 1534 & 730\\
        \hline
%        N$^{\mathrm{QMw}}_{\mathrm{frames}}$ & 70 & 60 & 20 & 20 & 20 & 20 & 20 & 20 & 20 & 20 & 20
    \end{tabular}
    }
    \caption{Number of exploited frames (N$_{\mathrm{frames}}$) and average number of solvent molecules (N$_{\mathrm{H_2O}}$) included in the droplets used in QM/MM calculations of each considered systems. }%Number of exploited frames (N$^{\mathrm{QMw}}_{\mathrm{frames}}$) in the QM/QMw/FQF$\mu$ calculations of each considered system.}
    \label{tab:geometries}
    \end{ruledtabular}
    \end{table}
\noindent

All QM/MM calculations are performed by using a locally modified version of ADF \cite{adf,ADF2001} engine within the Amsterdam Modeling Suite (AMS). \cite{AmsterdamModellingSuite} The QM part is treated by exploiting the CAMY-B3LYP functional \cite{akinaga2008range,seth_range-separated_2012} combined with the TZ2P basis set. \cite{van_lenthe_optimized_2003} Solvent molecules within the MM region are described by means of the aforementioned classical force fields.
In QM electrostatic embedding calculations, the water molecules are described by the TIP3P force field. \cite{mark2001structure}  Three parameterizations are employed in QM/FQ calculations:  FQ$_1$ (from Ref. \citenum{rick1994dynamical}), FQ$_2$ (from Ref. \citenum{carnimeo2015analytical}) and FQ$_3$ (from Ref. \citenum{giovannini2019effective}). Three parameterizations are also employed for QM/DRF calculations: DRF$_1$, DRF$_2$ and DRF$_3$ (all from Ref. \citenum{jensen2003discreteDFT}). The parameterization presented in Ref. \citenum{giovannini2019polarizable} is exploited in QM/FQF$\mu$ and QMw calculations.

\section{Results}\label{sec:results}
In this section we examine vacuo-to-water solvatochromic shifts.
% in light of the different physico-chemical bases of the exploited polarizable QM/MM approaches. 
First, we discuss reference computed data in vacuo. Second, we compare computed solvatochromic shifts obtained with each embedding model with their experimental counterparts, by resorting to selected statistical estimators to quantitatively analyze the performance of the various models. Finally, we rationalize our findings in light of the different physico-chemical description provided by the employed approaches.
Remarkably, QM/MM results are commonly directly compared with experimental findings, and for this reason we primarily assess the quality of the embedding models by taking the experimental solvatochromic shift as a reference. However, it is worth noting that experimental shifts might be affected by a variety of physical effects (temperature, Franck-Condon broadening, repulsion, dispersion, ...) which are only partially included (or even absent) in our QM/MM modeling. Then, QM/MM results are also compared to QMw data, which may be seen as complementary to experimental measurements. However, it is worth remarking that such a reference is not optimal, because QMw provides a QM description of solute-solvent interactions, thus including Pauli repulsion and charge transfer effects, which are not taken into account by a purely QM/classical approach. 
 
%For a broader discussion about the phenomena responsible for the appearance of the solvatochromic shift and how these can be modelled, we refer the reader to more specific works such as Refs. \citenum{giovannini2018polarizable,marini2010solvatochromism}.

\subsection{Excitation energies in vacuo}\label{sec:energies in vacuo}
Experimental vacuum excitation energies are only available in the literature for {\acrolein} and {\pNA}. In order to obtain reference "experimental" values for the other compounds, we resort to an extrapolation procedure, through a linear fit of the experimentally available excitation energies measured in different solvents, as a function of the solvent polarity indicator $E^N_T$. The value of $E^N_T$ in gas phase is set to -0.111 according to \textit{Reichardt et al.}. \cite{reichardt1994solvatochromic,reichardt2011solvents} The resulting values are reported in Tab. \ref{tab:Exp energies}, together with the fitting $R^2$ coefficients (see Sections S5.2 -- S5.12 in the {\SM} for more details).

\begin{table}[!h]
    \centering
    \begin{ruledtabular}
    \begin{tabular}{l|l|l|l|r|c}
         & \multicolumn{2}{c|}{VAC} & WTR & Shift & R$^2$\\
        \cline{1-6}
        molecule & Calc & Exp & Exp & Exp &\\
        \hline
        \acrolein$_{n \to \pi^*}$ & 3.78 & 3.69 \cite{aidas2008performance} & 3.94\cite{aidas2008performance} & -0.25 & -- \\
        \acrolein$_{\pi \to \pi^*}$ & 6.46 & 6.41\cite{aidas2008performance} & 5.89\cite{aidas2008performance} & 0.52 & --\\
        {\pNA}& 4.34 & 4.24 \cite{MILLEFIORI197721} & 3.26 \cite{KOVALENKO2000312} & 0.98 & --\\
        {\quinolinium}& 2.06 & 1.98$^*$& 2.80 \cite{novaki1996solvatochromism} & -0.82 & 0.998\\
        {\phtalimide}& 4.00 & 3.50$^*$& 3.35 \cite{soujanya1992nature} & 0.15 & 0.537 \\
       {\bimane} & 3.99 & 3.54$^*$& 3.20\cite{politzer1989effects} & 0.34 & 0.999\\
       {\cyanine}& 2.76 & 2.64$^*$ & 2.78\cite{abdel2009solvatochromism} & -0.14 & 1.000\\
       {\doxo} & 2.97 & 2.50$^*$ & 2.49 \cite{zakerhamidi2014photo} & 0.01 & 0.196 \\
       {\betaine} & 1.61 & 1.18 \cite{reichardt2011solvents} & 2.74 \cite{reichardt2011solvents} & -1.56 & 1.000\\
        \bodipy & 3.15 & 2.46$^*$& 2.49\cite{arroyo2009smallest} & -0.03 & 0.437\\
       {\merocyanine} & 2.94 & 1.56$^*$& 2.80 \cite{abdel2005absorption,doi:10.1021/j100280a012} & -1.24& 0.960\\
       {\citidina} & 4.79 & 4.30$^*$ & 4.46\cite{martinez2016computing} & -0.16 & 1.000\\
    \end{tabular}
    \end{ruledtabular}
    \caption{Calculated (Calc) and experimental (Exp) excitation energies (eV) in vacuo (VAC) and aqueous solution (WAT). The experimental values reported with $*$ are extrapolated by a linear fit (R$^2$ coefficients reported in the last column). Experimental solvatochromic shifts (eV) are also given.}
    \label{tab:Exp energies}
\end{table}

Acrolein (see {\acrolein} in Fig. \ref{fig:molecules set}) is experimentally characterized by a dark $n \to \pi^*$  and a bright $\pi \to \pi^*$ transition, which are placed at 3.69 and 6.41 eV, respectively.\cite{aidas2008performance} %\cite{marenich2011practical,moskvin1966experimental,aidas2008performance,aquilante2003theoretical,brancato2006quantum,duchemin2018bethe,bistafa2016complete,giovannini2019quantum}. 
Our calculations in gas-phase are in very good agreement with such findings, reporting a first dark transition at 3.78 eV 
and a second bright one at 6.46 eV (see Tab. \ref{tab:Exp energies}).
In the following analysis, such transitions are named {\acrolein}$_{n \to \pi^*}$ and {\acrolein}$_{\pi \to \pi^*}$, respectively.
For para-nitroaniline (pNA, see {\pNA} in Fig. \ref{fig:molecules set}), $\pi \to \pi^*$ transition is analyzed,\cite{sok2011solvent,giovannini2019electronic} which is experimentally measured at 4.24 eV,\cite{MILLEFIORI197721} and is well reproduced by our calculations (4.34 eV). 
The same occurs for 1-methyl-8-oxyquinolinium betaine (see {\quinolinium} in Fig. \ref{fig:molecules set}) for which our computed value is 2.06 eV, in good agreement with the extrapolated experimental value of 1.98 eV (see Sec. S5.4 in the {\SM}).

The experimental excitation energy of 4-aminophtalimide (see {\phtalimide} in Fig. \ref{fig:molecules set}) is extrapolated at 3.50 eV (see Sec. S5.5 in the {\SM}). Our computed value is placed at about 4.00 eV, differing of almost 0.5 eV from the extrapolated experimental value. Such a discrepancy is probably due to the choice of functional/basis set which may be not ideal for such system. We remark however that when studying solvatochromic shifts, which are obtained as energy differences, such systematic errors should not affect much the results.\cite{loco2016qm}

The first transition of syn-CH$_3$-CH$_3$ (bimane, see {\bimane} in Fig. \ref{fig:molecules set}), is computed at about 3.99 eV 
and is experimentally extrapolated at 3.54 eV (see Sec. S5.6 in the {\SM}).

4-2-[4-(dimethylamino)phenyl]ethenyl-1-methylpyridinium (see {\cyanine} in Fig. \ref{fig:molecules set}) is characterized by a computed first transition in vacuo at 2.76 eV, in good agreement with the 
corresponding experimental extrapolated excitation energy at 2.64 eV (see Sec. S5.7 in the {\SM}).

The first transition of doxorubicin (see {\doxo} in Fig. \ref{fig:molecules set}) in gas-phase falls at 3.00 eV 
, which differs of almost 0.5 eV from the extrapolated experimental excitation  energy (see Sec. S5.8 in the {\SM}).
Reichardt's dye (see {\betaine} in Fig. \ref{fig:molecules set}) first transition is calculated at about 1.61 eV 
in good agreement with its experimental counterpart (1.175 eV, see Sec. S5.9 in the {\SM}).
By moving to BODIPY (see {\bodipy} in Fig. \ref{fig:molecules set}), we report its first transition in vacuo at 3.15 eV.  
Its experimental excitation energy in vacuo is extrapolated at 2.46 eV (see Sec. S5.10 in the {\SM}).
For Brooker’s merocyanine (see {\merocyanine} in Fig. \ref{fig:molecules set}), the first transition has a $\pi \to \pi^*$ character and its maximum is placed at 2.94 eV, which largely differs from the extrapolated experimental excitation energy (1.56 eV, see Sec. S5.11 in the {\SM}). Note however that the computed value is in agreement with previously reported values (2.92 eV from Ref. \citenum{ambrosetti_quantum_2021}).

Finally, we studied the first transition of 5-methylcytidine (see {\citidina} in Fig. \ref{fig:molecules set}) which is computed at 4.79 eV. Its experimental excitation energy in vacuo is extrapolated at 4.30 eV (see Sec. S5.12 in the {\SM}).

\subsection{Solvatochromic Shifts}\label{sec:shift1}

\begin{figure*}[!htbp]
    \centering
    \includegraphics[width=0.9\textwidth]{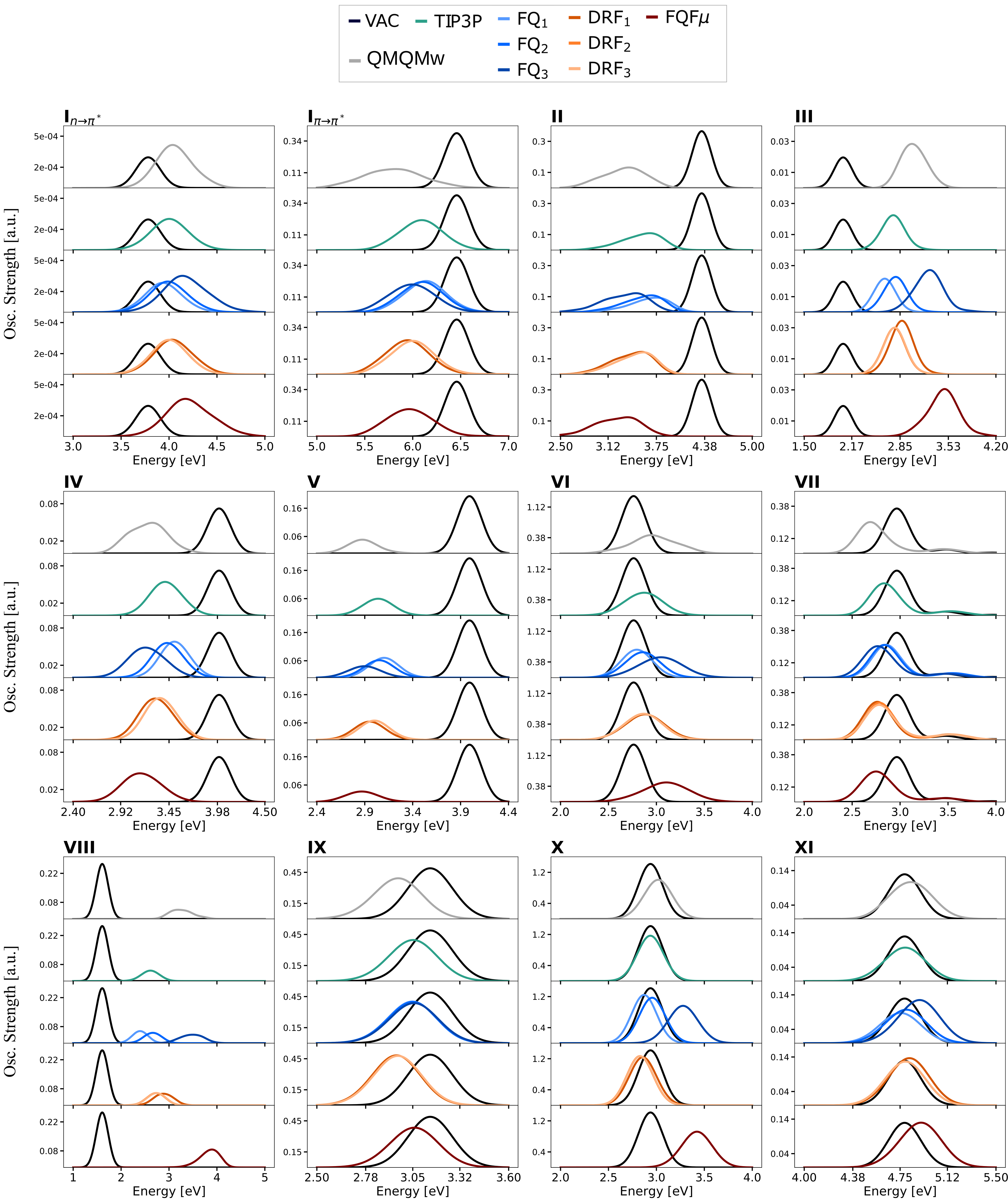}
    \caption{Computed absorption spectra of molecules {\acrolein}--{\citidina} in gas-phase (black) and in aqueous solution, as predicted by different embedding models.}
    \label{fig:spettri-tot}
\end{figure*}

\begin{figure*}[!htbp]
    \centering
    \includegraphics[width = 0.9\textwidth]{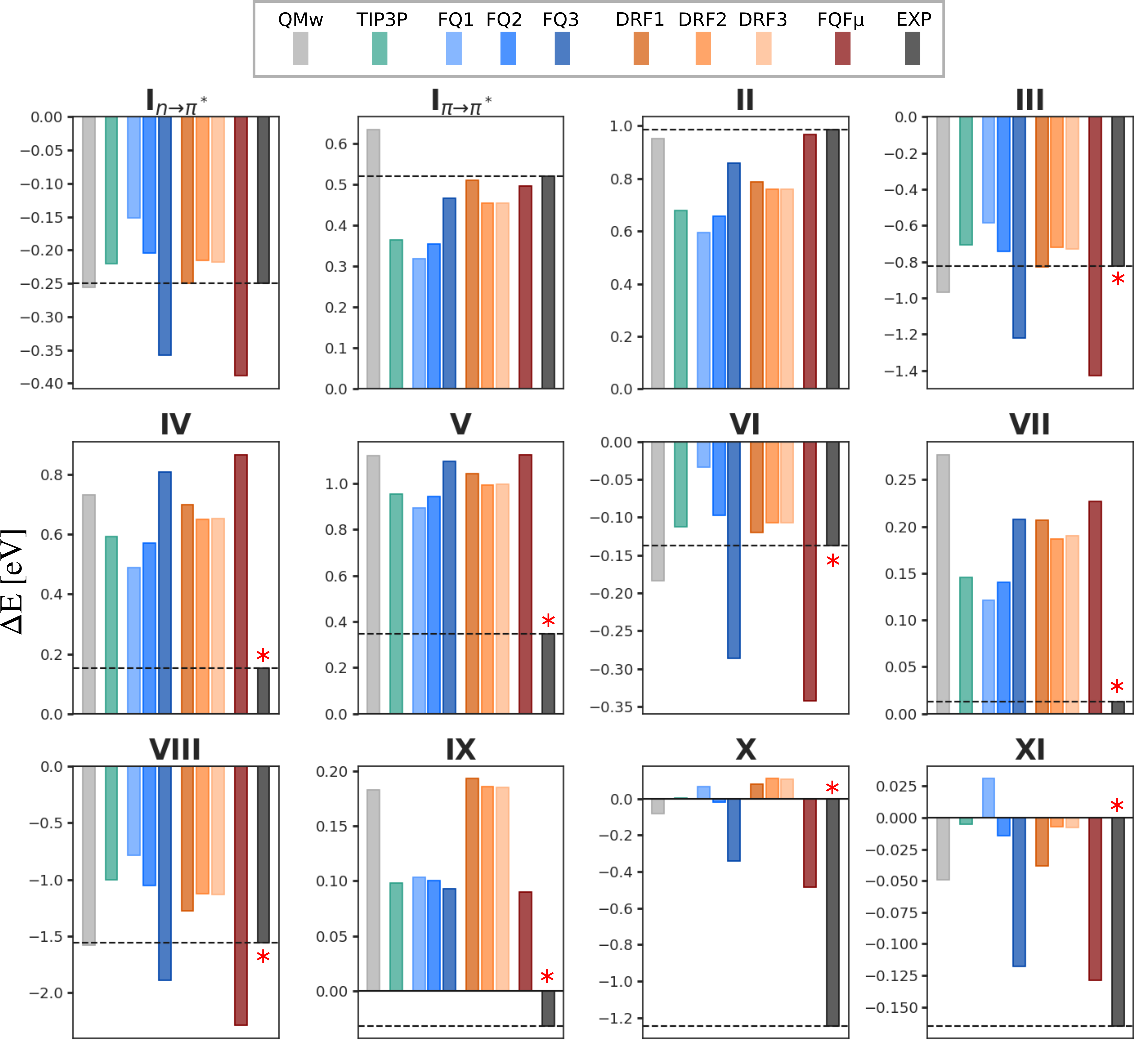}
    \caption{Computed solvatochromic shifts ($\Delta E$, in eV) with different QM/MM methods (see key). Experimental reference values are also plotted. Red asterisk denotes that the reference experimental gas-phase excitation energy has been extrapolated by a linear fit (see text).
    % In case of acrolein, two transitions {\acrolein}$_{n \to \pi^*}$ and {\acrolein}$_{\pi \to \pi^*}$ are considered (see Section S5.2 in the SM).
    }
    \label{fig:shifts comparison}
\end{figure*}

In Fig. \ref{fig:spettri-tot}, computed spectra of {\acrolein}--{\citidina} (see Fig. \ref{fig:molecules set}) in gas-phase and aqueous solution are reported. Raw data and pictures of the MOs involved in the transitions are shown in Sections S5.2 -- S5.12 in the {\SM}. The data depicted in Fig. \ref{fig:spettri-tot} clearly show that vertical excitation energies, oscillator strengths and band-widths, vary as a function of the specific embedding approach (and parametrization) which is exploited.
To rationalize such findings, we first investigate vacuo-to-water solvatochromic shifts ($\Delta E$) for {\acrolein} -- {\betaine}, as computed by exploiting the different embedding approaches discussed above. The results are graphically depicted in panels {\acrolein} -- {\betaine} in Fig. \ref{fig:shifts comparison}  (see Tabs. S4-S18 in the {\SM} for the raw data). 

All QM/MM approaches are able to qualitatively grasp experimental solvatochromic shifts, independently of the nature of the shift, i.e. batochromic ({\acrolein}$_\text{b}$, {\pNA}, {\phtalimide}, {\bimane}, {\doxo}) or hypsochromic ({\acrolein}$_{\text{a}}$, {\quinolinium}, {\cyanine}, {\betaine}). It is worth noting that the studied molecules display large variability in molecular size and, more importantly, amplitude of the solvatochromic shifts, which range from 0.01 eV ({\doxo}) to -1.56 eV ({\betaine}). 

We now focus on compounds {\bodipy}, {\merocyanine}, {\citidina} (see Tabs. S19-S24 in the {\SM} for the corresponding raw data). For {\bodipy}, all embedding approaches and the reference QMw method report a slight bathochromic shift. This is not in agreement with the small hypsochromic shift of -0.032 eV expected from our extrapolated linear fit. However, it is worth noting that, as previously reported, \cite{momeni_why_2015} TD-DFT fails to reproduce excitation energies of such a system. Therefore, the systematic failure of all embedding models to reproduce even the qualitative nature of the solvatochromic shift for {\bodipy} may be probably due to such an incorrect description of the electronic transition provided by TD-DFT. 
Moving to {\merocyanine}, the experimental hypsochromic shift is correctly described by FQ$_2$, FQ$_3$ and FQF$\mu$, but not by the other approaches (see Sec. S5.11 on the {\SM}). 
Remarkably, FQ$_2$, FQ$_3$ and FQF$\mu$ are able to grasp the sign of the shift, although the absolute value is underestimated, of about 1.23 eV (FQ$_2$), 0.91 (FQ$_3$) and 0.760 eV (FQF$\mu$). TIP3P, FQ$_1$ and all DRF parametrizations incorrectly predict the solvatochromic shift, both in absolute value and sign. A similar trend is also reported for {\citidina}: FQ$_3$ and FQF$_\mu$ perform much better than the other embedding approaches, by underestimating the fitted shift of 0.047 eV and 0.036 eV, respectively (see Sec. S5.12 on the {\SM}). In this case, only FQ$_1$ incorrectly describes the absolute value and sign of the experimental shift.

\subsection{Discussion}\label{sec:further}

The quality of the results reported in the previous section, and therefore the reliability and accuracy of the different embedding approaches, can be further quantified by means of the following statistical estimators (SE): 

\begin{align}
    \text{Mean Absolute Error (MAE)} & = \frac{\sum_i |x_i- \bar{x}_i|}{N} \\ 
    \text{Mean Error (ME)} & = \frac{\sum_i x_i - \bar{x}_i}{N} \\ 
    \text{Mean Error unsigned (ME$_u$)} & = \frac{\sum_i |x_i|-|\bar{x}_i|}{N} \\ 
    \text{Max ME$_u$ (MME$_u$)} & = \text{max} \left( |x_i|-|\bar{x}_i|  \right)
\end{align}

where $x_i$ and $\bar{x}_i$ are the computed and the reference solvatochromic shifts (either the experimental values or the computed QMw ones) of the $i$-th transition (see Fig. \ref{fig:shifts comparison}), respectively, whereas $N$ is the total number of transitions considered. 

In the following, we report computed SE values for the whole set of investigated systems by taking as reference experimental shifts (see Tab. \ref{tab:rmse_tot_exp}) and calculated QMw data (see Tab. \ref{tab:rmse_tot_QM/QMw}).

\begin{table}[!htbp]
    \centering
    \begin{ruledtabular}
    \begin{tabular}{l|ccccccccc}
         & QMw & TIP3P   & FQ$_1$ & FQ$_2$ & FQ$_3$ &  DRF$_1$ & DRF$_2$  & DRF$_3$ & FQF$\mu$\\
        \hline
        MAE  & 0.29 & 0.33 & 0.37 & 0.32 & 0.32 & 0.30 & 0.33 & 0.33 & 0.36\\
        ME   & 0.25 & 0.25 & 0.27 & 0.24 & 0.13 & 0.27 & 0.28 & 0.28 & 0.08\\
        ME$_u$ & 0.07 & -0.11 & -0.17 & -0.11 & 0.13 & -0.02 & -0.06 & -0.06 & 0.22\\
        MME$_u$  & -1.16 & -1.24 & -1.18 & -1.23 & -0.91 & -1.17 & -1.14 & -1.14 & 0.78\\
        succ \% & 92\% & 83\% & 75\% & 92\% & 92\% & 83\% & 83\% & 83\% & 92\% \\
    \end{tabular}
    \end{ruledtabular}
    \caption{Calculated MAE, ME, ME$_u$, MME$_u$ (eV) for all systems, as described by the various embedding approaches. Reported SE are computed with respect to reference experimental shifts. The success rate in percentage (succ \%) of each model in predicting the correct experimental sign of the shift is also reported.}
    \label{tab:rmse_tot_exp}
\end{table}

\begin{table}[!htbp]
    \centering
    \begin{ruledtabular}
    \begin{tabular}{l|cccccccc}
        & TIP3P & FQ$_1$ & FQ$_2$ & FQ$_3$ &  DRF$_1$ & DRF$_2$  & DRF$_3$ & FQF$\mu$\\
        \hline
        MAE$^{\mathrm{Q}}$   & 0.18 & 0.25 & 0.18 & 0.13 & 0.10 & 0.14 & 0.14 & 0.20\\
        ME$^{\mathrm{Q}}$ &  0.00 & 0.02 & -0.01 & -0.12 & 0.02 & 0.03 & 0.03 & -0.17\\
        ME$_{u}^{\mathrm{Q}}$ &  -0.18 & -0.24 & -0.18 & 0.06 & -0.08 & -0.12 & -0.12 & 0.15\\
        MME$_{u}^{\mathrm{Q}}$ & -0.57 & -0.75 & -0.53 & 0.31 & -0.31 & -0.45 & -0.45 & 0.71\\
        succ \% &  92 \% & 83\%  & 100 \% & 100 \% & 92 \% & 92 \% & 92 \% & 100 \% \\
    \end{tabular}
    \end{ruledtabular}
    \caption{Calculated MAE, ME, ME$_u$, MME$_u$ (eV) for all systems, as described by the various embedding approaches. Reported SE are computed with respect to calculated QM/QMw/FQF$\mu$ shifts. The success rate in percentage (succ \%) of each model in reproducing the sign of the computed QM/QMw/FQF$\mu$ shift is also reported.}
    \label{tab:rmse_tot_QM/QMw}
\end{table}

\noindent

By first focusing on Tab. \ref{tab:rmse_tot_exp}, we first note that MAE values range from 0.30 eV (DRF$_1$) to 0.37 eV (FQ$_1$) with 0.29 eV for QMw, indicating that all models similarly behave as far as absolute errors are considered. 
ME values confirm the best performance of FQ$_3$ and FQF$\mu$. However, low ME values may also reveal error cancellations, i.e. compensation of over- and underestimated data. In line with QMw results, FQF$\mu$ and FQ$_3$ are the only that predict positive ME$_u$ values: in particular the largest ME$_u$ value is reported for FQF$\mu$ and both models overestimate experimental solvatochromic shifts. The opposite is instead valid for all other embedding approaches, which tend to underestimate reference values. Among all methods, the lowest ME$_u$ values are associated to the DRF family: in particular for DRF$_1$ ME$_u = 0.02$ eV, i.e. lower than the chemical accuracy (1 kcal/mol). Low ME$_u$ values are certainly related to a good average reproduction of the experimental solvatochromic shifts. However, the MME$_u$ estimator reveals that large maximum errors can be reported by all methods, including DRF. In this case, FQF$\mu$ (followed by FQ$_3$) gives the lowest MME$_u$. Moreover, it is worth noting that only FQF$\mu$, FQ$_2$, and FQ$_3$ can correctly predict the sign of the experimental solvatochromic shifts for all molecules (except for {\bodipy}), reporting the same success rate of QMw calculations (see succ \% in Tab. \ref{tab:rmse_tot_exp}). 
To further rationalize our findings, we now discuss the SE computed by taking as a reference the QMw shifts (indicated with the superscript ``Q'', see Tab.  \ref{tab:rmse_tot_QM/QMw}). As stated above, QMw accounts for effects which are not considered in our QM/MM calculations (mainly charge transfer between solute and solvent molecules, and Pauli repulsion). 
We first note that MAE$^{\mathrm{Q}}$ is bound to lower values, ranging from 0.10 eV (DRF$_1$) to 0.25 eV (FQ$_1$), as compared to MAE (see Tab. \ref{tab:rmse_tot_exp}). The lowest ME$_u^{\mathrm{Q}}$ are reported for FQ$_3$ with 0.06 eV and DRF$_1$ with -0.08 eV, thus highlighting that, on average, FQ$_3$ overestimates the reference QMw shifts, whereas DRF$_1$ underestimates them. Notably, as shown by some of the present authors in Ref. \citenum{giovannini2019quantum}, the inclusion of Pauli repulsion in polarizable QM/MM calculations, might reduce, in absolute value, the computed solvatochromic shift and thus may compensate possible overestimation of the shifts with respect to QMw calculations. For this reason, we can argue that the inclusion of Pauli repulsion in QM/MM calculations might improve FQ$_3$ and FQF$\mu$ performance and worsen those of all other models. Indeed, this can be justified by the fact that FQ$_3$ and FQF$\mu$ parametrizations aim to reproduce electrostatic and polarization solute-solvent interactions only.\cite{giovannini2019effective,giovannini2019polarizable} Therefore, overestimation, in absolute value, of solvatochromic shifts is expected. Finally, an overall reduction of the MME$^Q_u$ indicator compared to the values in Tab. \ref{tab:rmse_tot_exp} is obtained, being the lowest values reported once again for FQ$_3$ (0.31 eV) and DRF$_1$ (-0.31 eV). 

\begin{figure*}[!htbp]
    \centering
    \includegraphics[width=.8\textwidth]{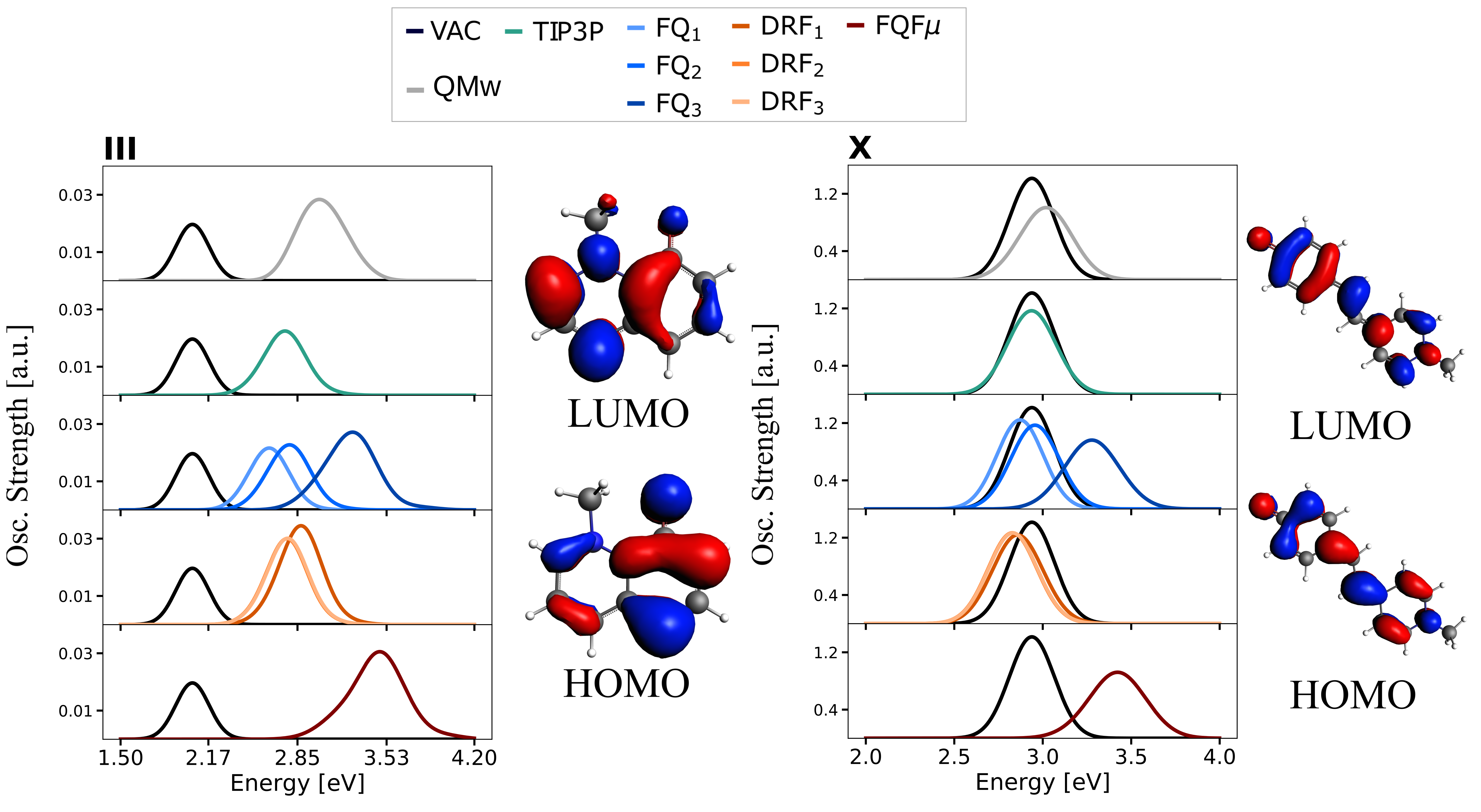}
    \caption{Calculated absorption spectra of {\quinolinium} (left) and {\merocyanine} (right) by exploiting the different QM/MM approaches. The calculated spectrum in gas-phase (black) and the corresponding HOMO and LUMO orbitals are also graphically depicted.}
    \label{fig:quinolinium-merocyanine}
\end{figure*}

% Lowest ME$_u^{\mathrm{Q}}$ are reported for FQ$_3$, which on average overestimates QMw computed shifts of 0.06 eV, and DRF$_1$ which on average underestimates the QMw computed shift of 0.08 eV.
The results discussed in terms of the statistical estimators can be better rationalized by focusing on two specific systems, i.e. quinolinium ({\quinolinium}) and Brooker's merocyanine ({\merocyanine}), characterized by an experimental hypsochromic shift. Computed absorption spectra, together with the main MOs involved in the transition (HOMO and LUMO), are reported in Fig. \ref{fig:quinolinium-merocyanine} left panel for molecule {\quinolinium} and right panel for molecule {\merocyanine}. The reported hypsochromic shift in polar solvents \cite{reichardt2004pyridinium} of {\quinolinium} is related to larger stabilization of the dipolar charge-separated ground state with respect to the non-zwitterionic excited state \cite{murugan2011modeling,KUMOI1970319}. A similar explanation has also been reported to explain the large hypsochromic shift of {\merocyanine} when dissolved in polar solvents\cite{wada2014solvatochromic,morley1997fundamental,wada2014solvatochromic,abdel2005absorption,tanaka2017solvatochromism}.

The left panel of Fig. \ref{fig:quinolinium-merocyanine} shows that all embedding approaches correctly predict the hypsochromic shift, with an associated enhancement of the absorption signal. In particular, the larger the computed solvatochromic shift, the higher enhancement is reported. Oppositely, only FQ$_2$, FQ$_3$ and FQF$\mu$ correctly reproduce the correct sign of the experimental shift of {\merocyanine}, also reporting a significant hypochromic effect (see Fig. \ref{fig:quinolinium-merocyanine} right panel). 
Such findings can be directly related to the physico-chemical differences of the embedding approaches. In fact, by first focusing on the FQ family, we recall that the FQ force field is defined in terms of $\chi$ and $\eta$, the former ($\chi$) being the source (together with the QM density) of charge redistribution, the latter ($\eta$) defining charge self-interaction. Thus, lower $\eta$ values and larger $\chi$ differences between different MM atoms imply that, at equilibrium, larger charges arise on the FQ atoms. As a consequence, the QM/MM electrostatic interaction is larger. Thus, larger distortion of the QM MOs, with an associated increased dipole moment, is expected.
The reported FQ trends can then be rationalized in terms of computed ground state dipole moments (see Tab. S25 in the {\SM}). 
In fact, FQ$_3$ enhances QM/MM site-specific interactions, possibly deforming GS orbitals more than FQ$_2$, and in turn FQ$_1$. 

By moving to the DRF family, we note that DRF$_2$ and DRF$_3$ give almost the same results, which are in agreement with previous studies \cite{jensen2003discreteDFT}. DRF$_1$ reports slightly larger solvatochromic shift, because it exploits lower oxygen polarizability ($\alpha$) and higher hydrogen $\alpha$. Then, the electrostatic QM/MM interaction is overall increased (see Tab. S26 in the {\SM}), thus resulting in higher shifts with respect to other DRF parametrizations. Nevertheless, estimated vacuo-to-water shifts show only minor differences among different DRF models. This is in line with Ref. \citenum{jensen2003discreteDFT}, and confirms that the dissection of water polarizability into atomic contributions does not affect much computed transition energies. Thus, the values of the fixed charges assigned to DRF atoms crucially determine vertical transition energies. 

FQF$\mu$ parametrization is characterized by lower oxygen and hydrogen polarizabilities as compared to DRF$_1$. However, FQF$\mu$ also considers polarizable charges. As a consequence, it provides the most intense QM/MM interaction, as confirmed by computed dipole moments (see Tab. S25 and Tab. S26 in the {\SM}). It is also worth noting that stronger QM/MM interaction is also reflected in a larger computed full-width-at-half-maximum (FWHM), which is directly related to a broader distribution of excitation energies (see also Sections S5.2 -- S5.12 in the {\SM}). The discussed trends can also be rationalized in terms of the average dipole moment of water in the liquid phase as modeled by exploiting the various embedding approaches (see Sec. S5.14 and Tab. S26 in the {\SM}).

This analysis of molecules {\quinolinium} and {\merocyanine}, allows us to conclude that the contribution of charges substantially affects the electronic response. This is not surprising and has been previously reported in other contexts.\cite{giovannini2019polarizable,mao2017performance} This rationalizes the rather good performance of the non-polarizable TIP3P in terms of statistical estimators (see Tab. \ref{tab:rmse_tot_exp} and Tab. \ref{tab:rmse_tot_QM/QMw}). However, TIP3P yields the largest MME$_u$ and MME$_u^Q$, thus highlighting the importance of polarization effects. 

In order to obtain a physico-chemical rationalization of the discrepancies in the estimated blue-shifts for {\merocyanine} and {\citidina}, we can exploit a simple model to theoretically explain the measured hypsochromic effects. Let us assume that both GS and ES are a linear combination of two states $|A\rangle$  and $|B\rangle$, i.e. $|\psi\rangle = c^{\psi}_{A}|A\rangle + c^{\psi}_{B}|B\rangle$, where $|\psi\rangle$ is either GS or ES. 
Let us now assume that $|A\rangle$ is a charge separated state, whereas $|B\rangle$ is a state diffused over the molecule. We consider two different cases:
\begin{enumerate}
    \item In vacuo, $c^{GS}_A \gg c^{GS}_B$ and $c^{ES}_A \ll c^{ES}_B$. Thus, GS is charge separated, whereas ES is mostly diffused on the whole molecule. We then expect large GS dipole moment ($\mu_{GS}$) and transition charge transfer (CT) indices. For example, in vacuo $\mu_{GS} = 6.42$ Debye, $\Delta r = 1.96$ {\AA} for {\quinolinium}, and $\mu_{GS} = 14.47$ Debye, $\Delta r = 4.48$ {\AA} for {\betaine}. 
    % CHIARA: QUI CT!!!! 
    When the molecule is dissolved in water, GS charge separation, together with $\mu_{GS}$, increases. Overall, electronic transitions of this type display a substantial blue shift ({\acrolein}$_{n \to \pi^*}$, {\quinolinium}, {\cyanine} and {\betaine}). 
        % See \cref{fig:deltar-blue-shifting-compounds} and table S25 in section S5.12 in the {\SM} for more details. 
    For these molecules, FQ$_3$ and FQF$\mu$ give the largest errors and consistently overestimate the shift. On the contrary, DRF$_1$ gives the best values and the other DRF models yield errors of $\sim$ 10 \% higher than DRF$_1$ for these transitions. 
    Therefore, embedding models that stabilize the GS are needed to grasp the blue-shifting nature of these transitions. However, models which are parameterized to provide strong QM/MM electrostatic interactions, such as FQ$_3$ and FQF$_\mu$, might lead to unphysical GS overstabilization. 
    \item In vacuo, $c^{GS}_A \approx c^{GS}_B$ and $c^{ES}_A \approx c^{ES}_B$. Thus, GS and ES are a superposition of both $|A\rangle$  and $|B\rangle$ states. Due to the similarities between GS and ES, we expect a small CT index in vacuo. 
    For example, in vacuo $\mu_{GS} = 15.60$ Debye, $\Delta r = 0.70$ {\AA} for {\merocyanine}, and $\mu_{GS} = 5.61$ Debye, $\Delta r = 0.64$ {\AA} for {\citidina}. 
    % CHIARA: DOVE SONO GLI INDICI CT???? 
    In aqueous solution, charge separation increases, and also $\mu_{GS}$ and $c^{GS}_A$. If the solvent removes the degeneracy between $|A\rangle$ and $|B\rangle$, GS and ES are expected to have substantially different charge distributions. In this case, the CT index is large, and the transition is blue-shifted  due to a strong GS stabilization. This is the case for {\merocyanine} (see Tab. S25 in the {\SM}). On the contrary, if both $c^{GS}_A$ and $c^{GS}_B$ increase in a similar manner, blue-shift is expected if $c^{GS}_A > c^{ES}_A$, i.e. if GS is more stabilized than ES. In this case, the solvatochromic shift is small and the CT index remains similar to its value in vacuo. A transition of this type is reported for {\citidina} (see Tab. S25 in the {\SM}). 
    For both transition types, FQF$\mu$ gives the lowest errors, followed by FQ$_3$. This suggests that, for this class of transitions, models that provide the strongest QM/MM interaction are needed to grasp the correct electronic reorganization of the states involved.
\end{enumerate}

\begin{figure}[!htbp]
    \centering
    \includegraphics[width=0.4\textwidth]{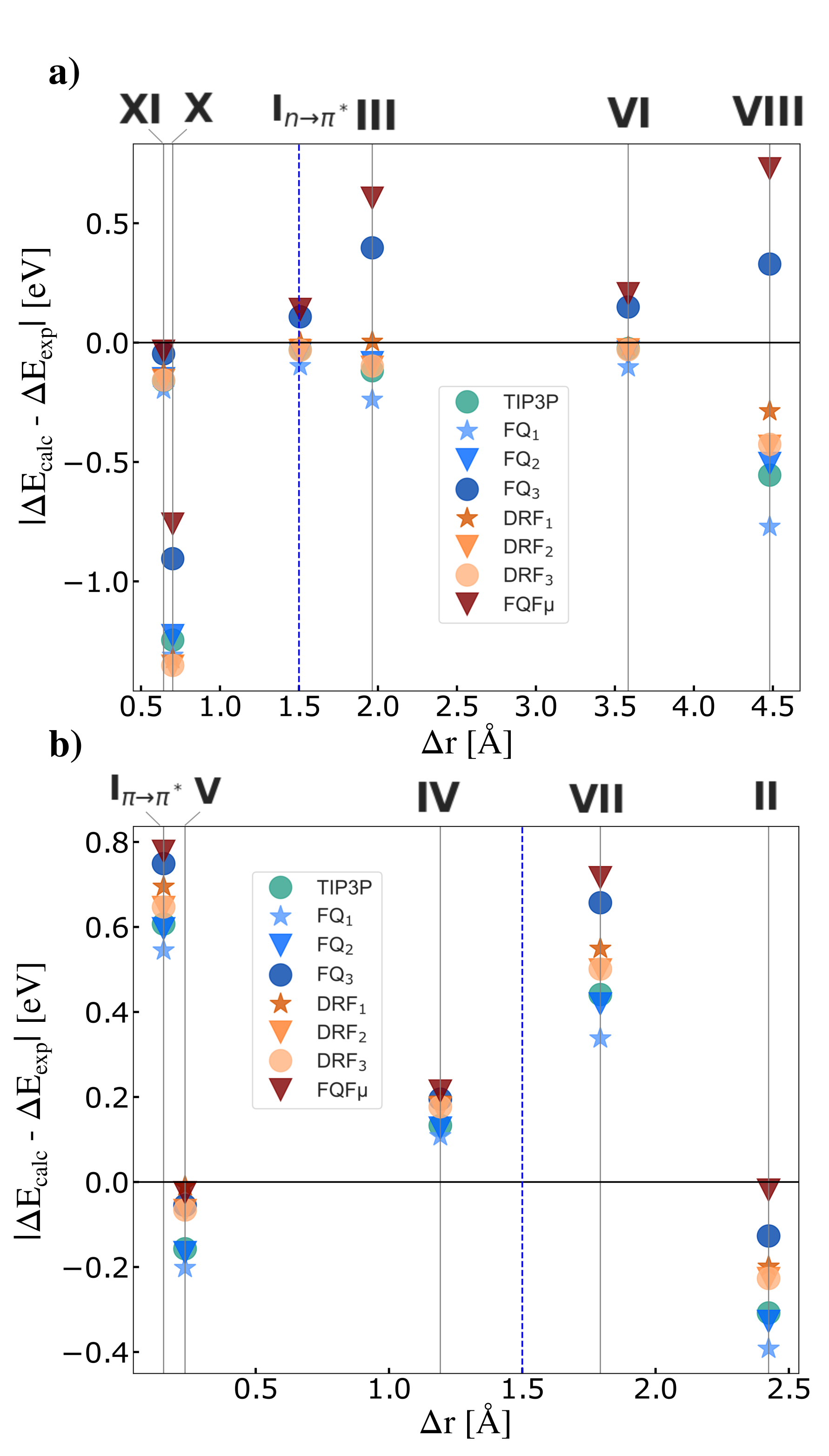}
    \caption{Correlation between gas-phase $\Delta r$ indices and absolute errors predicted by all embedding methods for blue-shifting (panel a) and red-shifting (panel b) transitions. Note that negative errors indicate underestimation of the shift with respect to reference experimental values.}
    \label{fig:delta-r-tot}
\end{figure}

Although qualitative, this analysis allows us to rationalize the behaviour of blue-shifting compounds in terms of the $\Delta r$-index calculated in vacuo. Indeed, large $\Delta r$-index calculated in vacuo (i.e. $\Delta r > 1.5$ {\AA}) tracks a transition with a strong CT character \cite{guido2013metric}, thus indicating that GS and ES have substantial different charge distributions. In Fig. \ref{fig:delta-r-tot} the absolute deviation of the computed shift from the estimated shift ($|\Delta E^{calc} - \Delta E^{exp}|$) is plotted with respect to calculated $\Delta r$-index in vacuo for each embedding model. Absolute errors less than zero indicate that the shift is underestimated with respect to the experiment.  In Fig. \ref{fig:delta-r-tot} a) the results obtained for the blue shifting compounds are depicted. If gas-phase $\Delta r < 1.5$ {\AA}, all models underestimate the shifts but FQF$\mu$ and FQ$_3$ report the lowest absolute errors. On the other hand, if $\Delta r > 1.5$ {\AA}, FQF$\mu$ and FQ$_3$ consistently overestimate the shifts, whilst FQ$_1$ reports the largest underestimation for each transition. In this case, DRF$_1$ reports the lowest absolute errors. In Fig. \ref{fig:delta-r-tot} b), the results obtained for red shifting compounds are instead considered. Interestingly, no particular correlation between the absolute estimation error and the $\Delta r$-index can be found. We notice however that for all molecules, DRF models compute similar shifts, FQF$\mu$ and FQ$_3$ yield the largest shifts, whereas FQ$_1$ gives the smallest shifts.

Therefore, our results show that none of the tested embedding model outperforms the others, because the agreement with experiments is strongly dependent on the nature of the transition under consideration. This is clearly depicted in Fig. \ref{fig:delta-r-tot} a) and finds a theoretical explanation in the two-states model described above.

\noindent

\section{Conclusions}\label{sec:conclusions}

In this work we have applied diverse QM/MM approaches and related parametrizations to the calculation of vacuo-to-water solvatochromic shifts. The selected QM/MM methods are based on the non-polarizable TIP3P, and polarizable FQ, DRF, and FQF$\mu$ force fields. For each model we have considered various parameterizations and computed shifts have been compared to experimentally available data and to computed QM/QMw/FQF$\mu$ values. 

For {\acrolein}--{\betaine}, all models are able to grasp the correct nature of the shift, however reporting a very large MME$_u$. FQ$_3$ and FQF$\mu$ consistently overestimate the shift (except for {\acrolein}$_{\pi \to \pi^*}$ and {\pNA}), which is consistent with the fact that such models have been parameterized by accounting for solute-solvent electrostatic and polarization interactions only. For {\bodipy}, {\merocyanine} and {\citidina}, FQ$_3$ and FQF$\mu$ instead yield the best agreement with experimental values, and remarkably, for {\merocyanine}, only FQ$_2$, FQ$_3$ and FQF$\mu$ can describe the experimentally measured hypsochromism. Moreover, when compared with QMw computed shifts, FQ$_3$ and DRF$_1$, report the best performance in terms of statistical estimators. However, we argue that the inclusion of Pauli repulsion in QM/MM calculations would reduce in absolute value the computed shifts, thus improving FQ$_3$ and  FQF$\mu$ results and worsening those of all other models.\cite{giovannini2017general} 

The different trends have been analyzed in terms of the physico-chemical description provided by the employed QM/MM approaches, in terms of the predicted strength of QM/MM electrostatic interactions and QM dipole moments, and by means of a simplified two-state model for the blue-shifting transitions. %CHIARA QUI VOGLIAMO METTERE QUALCHE ALTRO DETTAGLIO?

The results of this study reveal that the correct description of solvatochromic shifts is a delicate task. Polarizable QM/MM approaches are nowadays becoming a golden-standard for condensed phase simulations, however, their use as a black-box may potentially yield completely wrong predictions  as for molecule {\merocyanine}. 
Indeed, computed values not only depend on the QM/MM approach which is exploited, but also on the specific parameterization and on the nature of the solute's electronic transition to be described.

\section*{Supplementary Material}

Theoretical aspects related to the QM/MM interaction integral; details on FQ model; details on DRF model; details on FQF$\mu$ model; convergence of QM/QMw/FQF$\mu$ excitation energy as a function of the number of frames; computed results for {\acrolein}--{\citidina}; computed properties for the blue-shifting compounds; average dipole moment for liquid water.

\begin{acknowledgments}
We gratefully acknowledge the Center for High Performance Computing (CHPC) at SNS for providing the computational infrastructure.
\end{acknowledgments}

\section*{Conflict of Interest}

The authors have no conflicts to disclose.

\section*{Data Availability}

The data that support the findings of this study are available from the corresponding authors upon reasonable request.

\section*{References}
%\nocite{*}
\bibliography{aipsamp}% Produces the bibliography via BibTeX.

\end{document}